\documentclass[aps,twocolumn,pra,superscriptaddress,showpacs,tightenlines]{revtex4}
\usepackage{amssymb}
\usepackage{amsmath}
\usepackage{graphicx}
\usepackage{epsfig}
\usepackage{subfigure}
\usepackage{amsfonts}

\begin{document}

\title{Electromagnetic manipulation for anti-Zeno effect \\
in an engineered quantum tunneling process}
\author{Lan Zhou}
\affiliation{Institute of Theoretical Physics, Chinese Academy of Sciences, Beijing,
100080,China}
\affiliation{Department of Physics, Hunan Normal University, Changsha 410081, China}
\author{F. M. Hu}
\affiliation{Department of Mathematics, Capital Normal University,
Beijing, 100037,China}
\author{Jing Lu}
\affiliation{Department of Physics, Hunan Normal University, Changsha 410081, China}
\author{C. P. \surname{Sun}}
\email{suncp@itp.ac.cn}
\homepage{http://www.itp.ac.cn/~suncp}
\affiliation{Institute of Theoretical Physics, Chinese Academy of Sciences, Beijing,
100080,China}

\begin{abstract}
We investigate an anti-Zeno phenomenon as well as a quantum Zeno effect for
the irreversible quantum tunneling from a quantum dot to a ring array of
quantum dots. By modeling the total system with the Anderson-Fano-Lee model,
it is found that the transition from the quantum Zeno effect to quantum
anti-Zeno effect can happen by adjusting magnetic flux and gate voltage.
\end{abstract}

\pacs{03.65.Xp, 03.65.Ta, 73.63.Kv}
\maketitle

\section{Introduction}

The modern development of quantum technology enables people to control
quantum process of microscopic system by external field \cite%
{Lloyd,Scull,zanar, sun-xue, kuri1,kuri2}. In the point of view of quantum
mechanics, the objective of a quantum control is to reach a desired state
(called target state) from an initial state of the controlled system by
manipulating its external parameters. Some aspects in quantum information
can be understood according to quantum control \cite{Qc-qc}. For example,
quantum computation, which manipulates the evolution of a quantum system by
appropriate logic gate operations, is essentially a quantum control process
by external parameters. In quantum error correction, the feedback control is
used to detect the unwanted couplings and correct them \cite{Lloyd}. Quantum
measurement can also be regard as a special control process, which projects
the unknown state into the definite state that we desired with maximized
probability through wave function collapse.

In quantum control, an intriguing conception is to use the quantum Zeno effect%
\cite{Sudar,Misra}. Such effect freezes the evolution of a quantum state
through frequent measurements. For instance, in the quantum bang-bang control%
\cite{Lloyd}, the measurement operations are generalized by a sequence of
pulses. Recently a quantum control scheme associated with the effect
opposite to the quantum Zeno effect was discovered, which accelerate the
decay of the unstable state by frequent measurements. Such effect is called
anti-Zeno effect (AZE) \cite{Kurizki1,Lewens,Kurizki5,Levit,Kurizki7} or
inverse Zeno effect. This discovery opens a new area for quantum control and
has been used to control various physical systems, such as trapped atoms in
an optical-lattice potential \cite{Raizen}, a superconducting current-biased
Josephson junction \cite{Kurizki4}, ultracold atomic condensates \cite%
{Kurizki6}, and etc.

In this paper, we consider the anti-Zeno effect with an engineered system
formed by an experimentally accessible ring-type quantum dot array and an
extra quantum dot. Here, the extra dot is coupled with one of dot array.
Since it is an artificial system with more flexibly controlled parameters,
we can study the dynamic detail of the transition between quantum Zeno
effect and quantum anti-Zeno effect in the one-direction quantum tunneling
of electron from the extra dot to the quantum dot array. Our main purpose is
to find a way controlling the electron tunneling. Our investigation is
mainly based on the discovery that the $k$-space representation of the
quantum dot ring model is equivalent to the famous Anderson-Fano-Lee model
\cite{fano,ander,Lee}, which correctly describes the irreversible quantum
process of a single energy level coupling with a continuous-spectra bath.
Then the standard approach \cite{luisell} is used to obtain the analytic
solution for the quantum tunneling dynamics. We also consider the tunneling
dynamics of bosons in a one-dimensional optical lattice with the same
configuration as that of fermions.

This paper is organized as follows: In section \ref{sec:qd}, we describe the
engineered model of quantum dot array. Then we point out that its $k$-space
representation is essentially the Anderson-Fano-Lee model. In section \ref%
{sec:ed}, we study the quantum irreversible process of quantum tunneling in
the Heisenberg picture. In section \ref{sec:qt}, we calculate the modified
tunneling rate by successive projective measurements, which are performed on
one dot to detect whether an electron is trapped here. We also recur to a
numerical calculation to confirm our observation. In section \ref{sec:iq},
we discuss the similar problems for bosons. Finally in section \ref%
{sec:summary}, we conclude the paper with some remarks.

\section{\label{sec:qd}Quantum dot array model for one-direction quantum
tunneling}

\begin{figure}[tbp]
\includegraphics[bb=193 437 402 527, width=7 cm,clip]{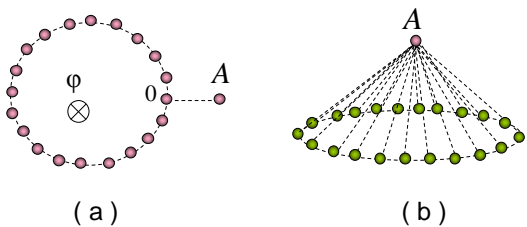}
\caption{\textit{(Color on line) (a)The real space schematic
illustration for 2N identical quantum dots arranged in a ring
threaded by a magnetic flux, with 0th dot interacting with dot $A$.
(b) The virtual space schematic illustration for a ring quantum dot
array coupled with quantum dot $A$ homogeneously.}} \label{m-f}
\end{figure}

We begin with a system of $2N$ identical quantum dots arranged in a ring
threaded by a magnetic flux $\phi $. Here, each structureless quantum dot
only traps one electron in a single state. The sites of the quantum dot ring
are labeled by $0,1$ $\cdots $ $2N-1$. The $0$th quantum dot interacts with
an additional quantum dot beside those placed on the ring, as is illustrated
schematically in Fig. \ref{m-f}(a).
%%%%%%%%%%%%%%%%%%%%%%%%%%%%%%%%%%%%%%%%%%%%%%%%%%%%%%%%%%%%%%%%%%%%
%%%%%%%%%%%%%%%%%%%%%%%%%%%%%%%%%%%%%%%%%%%%%%%%%%%%%%%%%%%%%%%%%%%%

Under the tight-binding approximation, the model Hamiltonian reads \cite%
{Peier,suncp2,Koski,Fath}
\begin{eqnarray}
H &=&\hbar J\sum_{j=0}^{2N-1}e^{i\frac{\pi }{N}\phi }\hat{a}_{j}^{\dagger }%
\hat{a}_{j+1}  \label{m-1} \\
&&+\hbar \omega _{A}\hat{a}_{A}^{\dagger }\hat{a}_{A}+\hbar g\hat{a}%
_{0}^{\dagger }\hat{a}_{A}+H.c.,  \notag
\end{eqnarray}%
which describes the electron tunneling dynamics of this quantum dots system
controlled by a magnetic flux. Here, $J$ denotes the hopping integral over
the $j$th site and the $j+1$th site. For simplicity, we assume $J$ is a
constant. $g$ is the coupling strength between two quantum dots at the $0$th
site and the additional site $A$; $\omega _{A}$ is the on-site potential (or
called the chemical potential) of $A$ site; $\phi $ is the magnetic flux
through the ring, and $a_{j}^{\dag }$($a_{j}$) is the fermion creation
(annihilation) operator at the $j$th site. We note here that the above
Hamiltonian was presented by Peierls \cite{Peier} to study the magnetic flux
effect phenomenologically up to the second order approximation.

We consider a dual picture (Fig.\ref{m-f}b) of the above quantum dot model
illustrated by Fig.\ref{m-f}(a). Through the Fourier transformation
\begin{equation}
\hat{a}_{j}=\frac{1}{\sqrt{2N}}\sum_{k=0}^{2N-1}e^{i\frac{\pi }{N}kj}\hat{a}%
_{k},  \label{m-2}
\end{equation}%
the original Hamiltonian is transformed into a $k$-space representation\cite%
{lieb}. In this momentum representation, the Hamiltonian becomes
\begin{eqnarray}
H &=&\hbar \sum_{k=0}^{2N-1}\epsilon _{k}\hat{a}_{k}^{\dagger }\hat{a}%
_{k}+\hbar \omega _{A}\hat{a}_{A}^{\dagger }\hat{a}_{A}  \label{m-3} \\
&&+\frac{\hbar g}{\sqrt{2N}}\sum_{k=0}^{2N-1}\left( \hat{a}_{k}^{\dagger }%
\hat{a}_{A}+h.c.\right) ,  \notag
\end{eqnarray}%
where
\begin{equation}
\epsilon _{k}=2J\cos \frac{\pi }{N}\left( \phi +k\right)   \label{m-4}
\end{equation}%
is the well-known Bloch dispersion relation. In this dual model
(\ref{m-3}), the quantum dots in a ring type array are coupled to
the single quantum dot $A $ homogeneously. The $2N$ modes of the
ring quantum dot array are characterized by the operators
$\hat{a}_{k}^{\dagger }$s and $\hat{a}_{k}$s, which create and
annihilate a quasi-excitation in the $k$th mode.

From the above dual picture of the quantum dot array model, it can
be observed that a one-direction quantum tunneling in our quantum
dot model can occur as a typical quantum dissipation phenomenon.
Since the quantum dot $A$ is coupled to other quantum dots of the
ring array, the electron in this dot can easily tunnel into the
array, but it is very difficult for all the electrons in the array
to go back to the dot $A$ simultaneously. Thus the electron in the
quantum dot $A$ will experience an irreversible process. This
similar phenomenon was studied as the Fano model \cite{fano} for
atom physics, the Anderson model for condensed matter physics
\cite{ander} and even as the Lee model for particle physics
\cite{Lee}. In this paper we focus on the quantum control problem
for the irreversible quantum tunneling, namely, we explore the
possibility of changing the microscopic quantum tunneling process by
adjusting the external fields, since many parameters in such an
artificially engineered system can be tuned to a great extent.

\section{\label{sec:ed} Evolution dynamics in Heisenberg picture}

The total system described by Hamiltonian (\ref{m-3}) is isolated as a
closed system, but the electron in each dot, such as dot $A$, is an open
system. When the dynamics of the system we are interested is only the
quantum dot $A$, the quantum dot array can be regarded as an engineered
environment. In terminology of quantum open system approach, the Hamiltonian
(\ref{m-3}) describes a single level system interacting with an environment
\cite{suncp1}. Such an engineered environment is composed of an ensemble of $%
2N$ qubits. State $|1\rangle $ denotes one electron in the dot, and $%
|0\rangle $ denotes no electron in the dot. The unitary operator generated
by the Hamiltonian (\ref{m-3}) entangles the system with the environment.

Now we investigate dynamics of the model (\ref{m-3}) in the Heisenberg
picture. The Heisenberg equation driven by the Hamiltonian (\ref{m-3})
results in the following equations
\begin{equation}
\frac{d}{dt}\hat{a}_{k}\left( t\right) =-i\epsilon _{k}\hat{a}_{k}-\frac{ig}{%
\sqrt{2N}}\hat{a}_{A},  \label{s-1}
\end{equation}%
\begin{equation}
\frac{d}{dt}\hat{a}_{A}\left( t\right) =-i\omega _{A}\hat{a}%
_{A}-ig\sum_{k=0}^{2N-1}\frac{\hat{a}_{k}}{\sqrt{2N}}. \label{s-2}
\end{equation}%
The motions of $\hat{a}_{k}$ and $\hat{a}_{A}$ are coupled via the
coupling constant $g$. For the convenience in the following
discussions, we only consider its short-time behavior, by employing
the operator ordering prescription. We should point out that the
short-time behavior has been studied in Ref\cite{Kurizki1} for the
general case with the coupling of a
discrete state to a continuum. With an analytical approach in Schr\"{o}%
dinger picture, they found that the decay processes of the single state
coupling to a discrete or a continuous spectrum is determined by the energy
spread incurred by the measurements\cite{Kurizki1}. Our approach will be
carried out in the Heisenberg picture for the present realistic system.

Defining two new fermion operators
\begin{equation}
\hat{C}_{k}=\hat{a}_{k}e^{i\epsilon _{k}t},\hat{B}=\hat{a}_{A}e^{i\omega
_{A}t},  \label{III-2}
\end{equation}%
to remove the high frequency effect, we have the integral-differential
equation as
\begin{eqnarray}
\frac{d\hat{B}}{dt} &=&-\frac{ig}{\sqrt{2N}}\sum_{k=0}^{2N-1}e^{-i\left(
\epsilon _{k}-\omega _{A}\right) t}\hat{C}_{k}\left( 0\right)   \notag \\
&&-\frac{g^{2}}{2N}\sum_{k=0}^{2N-1}\int_{0}^{t}\hat{B}\left(
t_{1}\right) e^{i\left( \epsilon _{k}-\omega _{A}\right)
(t_{1}-t)}dt_{1}  \label{III-5}
\end{eqnarray}%
from the above Eqs.(\ref{s-1}) and (\ref{s-2}). Integrating both sides of Eq.(%
\ref{III-5}), we proceed with an iteration method to obtain the suitable
operator ordering prescription for the dynamic evolution of $\hat{a}%
_{k}\left( t\right) $ and $\hat{a}_{A}\left( t\right) $. If the coupling
strength $g$ is small, we can omit the terms with the order of $g$ higher
than two. It is a reasonable assumption that $\hat{a}_{A}\left( t\right) $
varies slowly within a short time interval. By replacing $\hat{B}\left(
t_{1}\right) $ with $\hat{B}\left( 0\right) $ in the right hand side of the
above equation, the evolution of annihilation operator $\hat{a}_{A}\left(
t\right) $ is approximately calculated as
\begin{eqnarray}
\hat{a}_{A}\left( t\right)  &=&\hat{a}_{A}\left( 0\right) e^{-i\omega _{A}t}-
\notag \\
&&\sum_{k=0}^{2N-1}\frac{ig\hat{a}_{k}\left( 0\right) }{\sqrt{2N}}%
e^{-i\omega _{A}t}\int_{0}^{t}e^{-i\left( \epsilon _{k}-\omega _{A}\right)
t^{\prime }}dt^{\prime }  \label{III-6} \\
&&-\hat{a}_{A}\left( 0\right) e^{-i\omega _{A}t}\int_{0}^{t}dt^{\prime
}\left( t-t^{\prime }\right) e^{i\omega _{A}t^{\prime }}\Phi \left(
-t^{\prime }\right) ,  \notag
\end{eqnarray}%
where the memory function \cite{Kurizki4}%
\begin{equation}
\Phi \left( t\right) =\frac{g^{2}}{2N}\sum_{k=0}^{2N-1}e^{i\epsilon _{k}t},
\end{equation}%
only depends on the quasi-excitation in the $2N$ modes of the ring quantum
dot array and the magnetic flux.

\section{\label{sec:qt}Quantum tunneling affected by a sequence of
projective measurements}

Now we consider the decay of tunneling rate induced by an
instantaneous projective measurement into the initial state of the
total system. Suppose that the entire system is initially prepared
in a state with an electron in the quantum dot $A$ and no electron
in the ring array. Let $\left\vert 0\right\rangle $ denotes the
vacuum state that no electron exists in the entire system. Then, the
initial state can be written as
\begin{equation}
\left\vert \psi \left( 0\right) \right\rangle =\hat{a}_{A}^{\dagger }\left(
0\right) \left\vert 0\right\rangle .  \label{IV-1}
\end{equation}%
Obviously, this state is unstable since the electron may tunnel to any dot
of the quantum dot array in Fig.\ref{m-f}(b). After a period of evolution,
the probability for finding the electron inside the dot $A$ and no electrons
in the ring array is
\begin{equation}
p\left( t\right) =\left\vert \left\langle \psi \left( 0\right) \right\vert
U\left( t\right) \left\vert \psi \left( 0\right) \right\rangle \right\vert
^{2},  \label{IV-2}
\end{equation}%
where $U\left( t\right) =\exp \left( -iHt/\hbar \right) $ is the unitary
operator.

Assume the coupling strength $g$ is small. For a projective measurement into
the initial state, the probability for finding the electron in the initial
state is
\begin{equation}
p\left( t\right) =\exp \left( -Rt\right) ,  \label{IV-3}
\end{equation}%
which decays exponentially with a decay rate $R$ calculated as
\begin{equation}
R=2\text{Re}\int_{0}^{t}dt^{\prime }G(t,t^{\prime })e^{-i\omega
_{A}t^{\prime }}\Phi \left( -t^{\prime }\right) ,  \label{R}
\end{equation}%
where $\Phi \left( t^{\prime }\right) $ is just the memory function we
defined above and $\ G(t,t^{\prime })=\left( 1-t^{\prime }/t\right) $. It
has the similar expression to what obtained in Refs.\cite%
{Kurizki1,Lewens,Kurizki5,Levit,Kurizki7}.

To justify the above result, we assume the system is initially prepared in $%
\left\vert \psi \left( 0\right) \right\rangle =\hat{a}_{A}^{\dagger }\left(
0\right) \left\vert 0\right\rangle $. The probability for finding the
electron inside the dot $A$ and no electrons in the array is $p\left(
t\right) =\left\vert \left\langle 0\right\vert \hat{a}_{A}\left( t\right)
\left\vert 1_{A}\right\rangle \right\vert ^{2}$. With the explicit
expression Eq.(\ref{III-6}) for $\hat{a}_{A}\left( t\right) $, we obtain
\begin{equation}
p\left( t\right) =\left\vert 1-\int_{0}^{t}dt^{\prime }\left( t-t^{\prime
}\right) e^{-i\omega _{A}t^{\prime }}\Phi \left( -t^{\prime }\right)
\right\vert ^{2}.
\end{equation}%
Since $g$ is small and $\Phi \left( t\right) $ is proportional to $g^{2}$,
we approximately have%
\begin{equation}
p\left( t\right) \approx \left\vert e^{-\int_{0}^{t}dt^{\prime }\left(
t-t^{\prime }\right) e^{-i\omega _{A}t^{\prime }}\Phi \left( -t^{\prime
}\right) }\right\vert ^{2}  \label{IV-4}
\end{equation}%
or Eq.(\ref{IV-3}) with $R(t)$ defined by Eq.(\ref{R}).

After such measurements have been done by $n=t/\tau $ times, the survival
probability for finding the electron still in dot $A$ is
\begin{eqnarray}
p\left( t=n\tau \right) &=&e^{-2n\int_{0}^{\tau }dt^{\prime }\left( \tau
-t^{\prime }\right) e^{-i\omega _{A}t^{\prime }}\Phi \left( -t^{\prime
}\right) }  \notag \\
&=&e^{-2t\int_{0}^{\tau }dt^{\prime }G(\tau ,t^{\prime })e^{-i\omega
_{A}t^{\prime }}\Phi \left( -t^{\prime }\right) } \\
&=&e^{-2t\text{Re}\int_{0}^{+\infty }dt^{\prime }G(\tau ,t^{\prime })\Theta
\left( \tau -t^{\prime }\right) e^{-i\omega _{A}t^{\prime }}\Phi \left(
-t^{\prime }\right) },  \notag
\end{eqnarray}%
which gives the decay rate modified by measurement
\begin{equation}
R=2\text{Re}\int_{0}^{+\infty }dt^{\prime }G(\tau ,t^{\prime })\Theta \left(
\tau -t^{\prime }\right) e^{-i\omega _{A}t^{\prime }}\Phi \left( -t^{\prime
}\right) ,
\end{equation}%
where $\Theta \left( x\right) $ is the Heaviside unit step function, i.e. $%
\Theta \left( x\right) =1$ for $x>0$, and $\Theta \left( x\right) =0$ for $%
x<0$.

Define the modulation function caused by measurement as
\begin{equation}
f\left( t\right) =G(\tau ,t)e^{-i\omega _{A}t}\Theta \left( \tau -t\right) .
\label{IV-5}
\end{equation}%
By applying the Fourier transformation to the modulation function $f\left(
t\right) $ and the memory function $\Phi \left( -t\right) $, the decay rate
modified by the projective frequently measurement is calculated as
\begin{equation}
R=\frac{g^{2}\tau }{4\pi N}\sum_{m=0}^{2N-1}sinc^{2}\left[ \left( J\cos
\frac{\left( \phi +m\right) \pi }{N}-\frac{\omega _{A}}{2}\right) \tau %
\right] .  \label{IV-8}
\end{equation}%
Eq.(\ref{IV-8}) shows that the decay rate $R$ depends on four parameters:
the time interval $\tau $ between two successive measurements; the number $%
2N $ of quantum dots placed on the ring; the on-site-potential $\omega _{A}$
which is applied to the dot $A$ by the electrode; and the magnetic flux $%
\phi $ through the ring quantum dot array, but only $\tau $, $\omega _{A}$
and $\phi $ can be adjusted experimentally.

\begin{figure}[ptbh]
\includegraphics[bb=42 262 541 677, width=8 cm,clip]{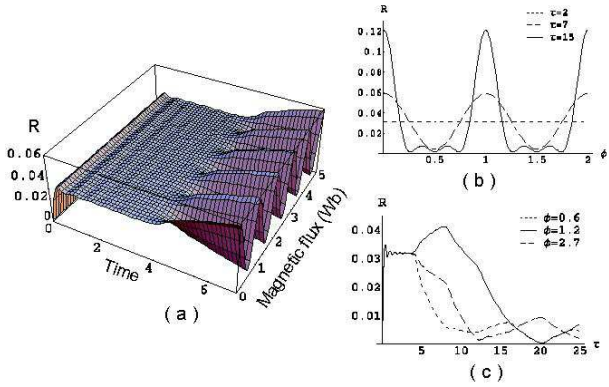}
\caption{\textit{(Color on line)(a) 3-D diagram for the behavior of
the decay rate as a function of $\protect\tau$ and $\protect\phi$
under the setting $J=5,g=1,N=20$ and $\protect\omega_{A}=0$. (b) The
cross sections of the 3-D surface for $\protect\tau=2, 7, 15$.
(c)The cross sections of the 3-D surface for $\protect\phi=0.6, 1.2,
2.7$. It shows the tunneling rate can be modulated by the magnetic
flux. The unit of time interval is $\hbar$ and the unit of magnetic
flux is Wb.}} \label{fw-tphi}
\end{figure}

To study the dynamic details of the irreversible quantum tunneling, we first
consider the dynamic behavior of electron with no measurement performed.
Fermi golden rule is used to calculate the decay rate as
\begin{equation}
R=\frac{g^{2}}{4N}\sum_{m=0}^{2N-1}\delta \left( 2J\cos \frac{\left( \phi
+m\right) \pi }{N}-\omega _{A}\right) .  \label{IV-9}
\end{equation}%
Eq.(\ref{IV-9}) shows the decay rate depends on $\omega _{A}$, $N$ and $\phi
$. If $\left\vert \omega _{A}\right\vert \leq 2J$ and the number of quantum
dots placed on the ring is finite, two situations happen to the electron
motion when one adjusts the magnetic flux $\phi $: 1) The electron tunnels
into the quantum dot array arranged in the ring and never come back; 2) the
electron stays in site $A$. In the following, we will explain the physical
mechanism for the switch between these two situations by adjusting $\phi $:
the energy level of the ring quantum dot array is discrete in Fig.\ref{m-f}%
(b), and the electron tunneling between dots occurs when the discrete energy
level of one dot matches that of the other dot. The magnetic flux $\phi $
controls the discrete energy levels of the quantum dot array to match or
not to match the energy level of quantum dot $A$ so that the above two
phenomenon occur. As the number of quantum dots placed on the ring
increases, the discrete energy levels of the dot array approach with each
other. Thus the effect of magnetic flux $\phi $ becomes vanishing, and the
controllable parameter is only the on-site potential $\omega _{A}$. The two
phenomena described above happen to the electron when $\left\vert \omega
_{A}\right\vert $ is smaller or larger than $2J$.

Actually, as for Eq.(\ref{IV-9}), one can also use the Wigner-Weisskopf
approach \cite{luisell} to describe the electron dynamic evolution
approximately. To this end, we first take the Laplace transformation of Eq.(%
\ref{III-5})
\begin{equation}
\hat{B}\left( s\right) =\frac{\hat{B}\left( 0\right) }{f\left( s\right) }%
-\sum_{k=0}^{2N-1}\frac{ig\hat{C}_{k}\left( 0\right) }{\sqrt{2N}f\left(
s\right) \left[ s+i\left( \epsilon _{k}-\omega _{a}\right) \right] }\text{,}
\end{equation}%
where
\begin{equation}
f\left( s\right) =s+\sum_{k=0}^{2N-1}\frac{g^{2}/\left( 2N\right) }{%
s+i\left( \epsilon _{k}-\omega _{a}\right) }.
\end{equation}%
As the coupling strength $g$ is small, the Wigner-Weisskopf approach gives
the zero point of $f\left( s\right)$ \cite{luisell}, which results in the
approximate solution
\begin{equation}
\hat{B}\left( t\right) =\hat{B}\left( 0\right) e^{-Rt}-\sum_{k=0}^{2N-1}%
\frac{ige^{-Rt}\hat{C}_{k}\left( 0\right) }{\sqrt{2N}[i\left( \epsilon
_{k}-\omega _{A}\right) -R]}\text{,}
\end{equation}%
where $R$ has the same expression as Eq.(\ref{IV-9}). So long as the
Wigner-Weisskopf approximation is valid for some time interval, the above
solution can correctly describe the quantum tunneling phenomenon in the
coupling quantum dot configuration.

Next we study the dynamic behavior of electron in quantum tunneling when $%
\tau\rightarrow0$, i.e. the system is measured continuously. In this case
the decay rate for the electron tunneling from quantum dot $A$ to the
quantum dot array vanishes. This means the electron is frozen in the quantum
dot $A$.

\begin{figure}[ptbh]
\includegraphics[bb=138 152 456 678, width=5 cm,clip]{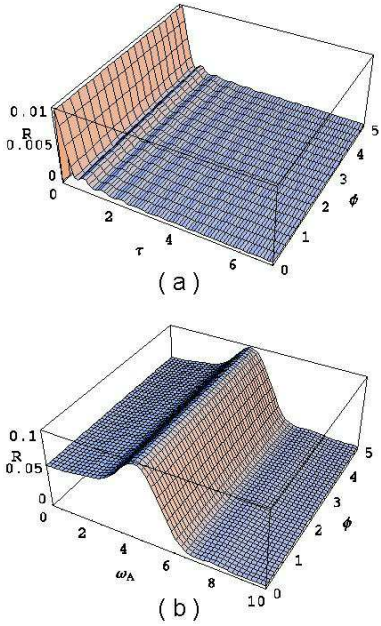}
\caption{\textit{(Color on line) The independence of the decay rate $R$ on
magnetic flux $\protect\phi$. (a) 3-D diagram for the behavior of the decay
rate as a function of $\protect\tau$ and $\protect\phi$ under the setting $%
J=5,g=1,N=20$ and $\protect\omega_{A}=20$. (b) 3-D diagram for the behavior
of the decay rate as a function of $\protect\omega_{A}$ and $\protect\phi$
under the setting $J=2.5,g=1,N=20$ and $\protect\tau=10$.}}
\label{r-phi}
\end{figure}

Then we consider the behavior of the electronic quantum tunneling with the
finite time interval between two successive measurements. Due to the
finiteness of time interval, we find that only quantum anti-Zeno effect can
occur in some cases. From Eq.(\ref{IV-8}), we can see when one of the energy
levels of the ring dot array matches that of dot $A$, that is, the
parameters $\phi $ and $\omega _{A}$ satisfy the following equation
\begin{equation}
2J\cos \frac{\left( \phi +m\right) \pi }{N}=\omega _{A}\text{,}
\label{IV-10}
\end{equation}%
the tunneling rate is an increasing function of time interval $\tau $.
Consequently, the quantum Zeno effect occurs. When all energy levels of the
array are out of resonance with that of dot $A$, i.e. Eq.(\ref{IV-10}) can
not be satisfied for any $m$, the tunneling rate is roughly a descending
function of $\tau $. Thus the quantum anti-Zeno effect occurs. Hence, when
the time interval $\tau $ between two successive measurements is finite, in
the region of $|\omega _{A}|<2J$, the occurrence of quantum Zeno or
anti-Zeno effect depends on the magnetic flux $\phi $ for a given on-site
potential $\omega _{A}$; and for a given magnetic flux $\phi $, the
occurrence of quantum Zeno or anti-Zeno effect depends on on-site potential $%
\omega _{A}$. In the region of $|\omega _{A}|>2J$, only quantum anti-Zeno
effect occurs when the time interval $\tau $ is in a finite appropriate
range.

In Fig.\ref{fw-tphi} and Fig.\ref{r-phi}(a), we numerically plot the decay
rate as the function of $\tau $ and the magnetic flux $\phi $ for different
on-site potential $\omega _{A}$. Fig.\ref{fw-tphi} is plotted when the
on-site potential $\omega _{A}$ is just within the energy range of the ring
dot array. It shows that, when the time interval approaching zero, the
quantum Zeno effect does occur, which coincides with our above discussion;
for a small time interval, the tunneling rate is a constant; for a
appropriate time interval, whether the electron tunnels out of the quantum
dot $A$ to other dot or stays in quantum dot $A$ is dependent on the
magnetic flux. This means we can inhibit or accelerate the dissipative
motion of the electron. Fig.\ref{r-phi}(a) shows that when the on-site
potential $\omega _{A}$ is outside of the energy range $\left[ -2J,2J\right]
$, the tunneling rate only depends on the interval $\tau $. As $\tau
\rightarrow 0$, the quantum Zeno effect also occurs, but there exists a
range of a finite $\tau $, in this range, the system decays rapidly as the
measurement frequency increases, so only the quantum anti-Zeno effect
occurs. These verify our above arguments.

\begin{figure}[tbph]
\includegraphics[bb=50 261 540 577, width=8 cm,clip]{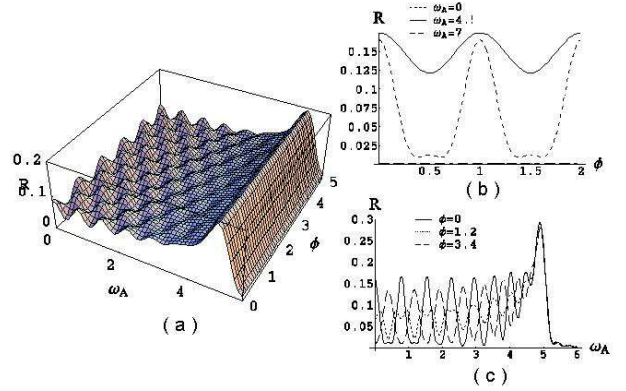}
\caption{\textit{(Color on line)(a) 3-D diagram for the behavior of the
decay rate as a function of $\protect\omega _{A}$ and $\protect\phi $ under
the setting $J=2.5,g=1,N=20$ and $\protect\tau =10$. (b) The cross sections
of the 3-D surface for $\protect\omega _{A}=0,4.5,7$. (c) The cross sections
of the 3-D surface for $\protect\phi =0,1.2,3.4$. It shows that the motion
of the electron can be modulated by electromagnetism.}}
\label{ft-wphi}
\end{figure}
For different time interval $\tau $ between two successive measurements, in
Fig.\ref{r-phi}(b) and Fig.\ref{ft-wphi}, we numerically plot the tunneling
rate as the function of the magnetic flux $\phi $ and on-site potential $%
\omega _{A}$. In this system, $\omega _{A}$ is controlled by the
electrochemical gate electrode. It can be seen that for sufficient small
interval $\tau $, shown in Fig.\ref{r-phi}(b), the tunneling rate modified
by measurement is independent of the magnetic flux $\phi $, but for an
appropriate interval $\tau $, shown in Fig.\ref{ft-wphi}, one can modulate
the tunneling rate by the magnetic flux when the on-site potential $\omega
_{A}$ is smaller then $2J$.

\section{\label{sec:iq}Irreversible quantum tunneling of boson in optical
lattice}

We consider the bosonic atoms trapped in a ring optical lattice\cite%
{Catali,Wang}, which is described as a periodic potential $V\left( x\right)
=V\left( x+a\right) $ with spatial period $a$. In general, we can use the
many-body Hamiltonian%
\begin{eqnarray}
H &=&\int \Psi ^{\dag }\left( x\right) \left[ \frac{p^{2}}{2m}+V\left(
x\right) \right] \Psi \left( x\right)  \\
&&+\int dxdy\Psi ^{\dag }\left( x\right) \Psi ^{\dag }\left( y\right)
W\left( x,y\right) \Psi \left( x\right) \Psi \left( y\right)   \notag
\end{eqnarray}%
to describe quantum dynamics of many-atom system. In the case of dilute
atomic gas, we can neglect the interaction term. When each potential well in
the optical lattice is deep sufficiently, the tight-binding approximation
can be used by assuming the wave function as $\Psi \left( x\right)
=\sum_{j}b_{j}u_{j}\left( x\right) $, where $u_{j}\left( x\right) $ is
localized around the site $j$. If we neglect the overlaps of two localized
basis states which are not next-neighbor, the coefficient $b_{j}$ will be
approximately described as a boson operator. Hence the Hamiltonian of such
boson system \cite{Zolle,Porto,Bhat,knight,Martin} can be approximated as
Eq.(\ref{m-1}) with $\phi =0$
\begin{eqnarray}
H &=&\hbar J\sum_{j=0}^{2N-1}\hat{b}_{j}^{\dagger }\hat{b}_{j+1}+\hbar
\omega _{A}\hat{b}_{A}^{\dagger }\hat{b}_{A}  \label{v-H} \\
&&+\hbar g\hat{b}_{0}^{\dagger }\hat{b}_{A}+H.c\text{.}  \notag
\end{eqnarray}%
Here, $\hat{b}_{j}^{\dag }$($\hat{b}_{j}$) is the creation (annihilation)
operator of bosonic atoms and they satisfy the commutation relations.

By the Fourier transformation for the boson operators $\hat{b}_{j}^{\dag }$
and $\hat{b}_{j}$, the boson model (\ref{v-H}) can be transformed into a
dual model similar to that of fermions (see the Eq.\ref{m-3})
\begin{eqnarray}
H &=&\sum_{k=0}^{2N-1}\varepsilon _{k}\hat{b}_{k}^{\dag }\hat{b}_{k}+\hbar
\omega _{A}\hat{b}_{A}^{\dag }\hat{b}_{A} \\
&&+\frac{g}{\sqrt{2N}}\sum_{k=0}^{2N-1}\left( \hat{b}_{k}^{\dagger }\hat{b}%
_{A}+\hat{b}_{A}^{\dagger }\hat{b}_{k}\right) ,  \notag
\end{eqnarray}%
where the Bloch dispersion relation is $\varepsilon _{k}=2J\cos \left( \pi
k/N\right) $.

We now use the Heisenberg equation to study the system dynamics. By
considering the short-time behavior which is described in section III, we
find the evolution of annihilation operator $\hat{b}_{A}$ is similar to Eq.(%
\ref{III-6})%
\begin{eqnarray}
\hat{b}_{A}\left( t\right)  &=&\hat{b}_{A}\left( 0\right) e^{-i\omega _{A}t}-
\label{v-op} \\
&&\sum_{k=0}^{2N-1}\frac{ig\hat{b}_{k}\left( 0\right) }{\sqrt{2N}}%
e^{-i\omega _{A}t}\int_{0}^{t}e^{-i\left( \varepsilon _{k}-\omega
_{A}\right) t^{\prime }}dt^{\prime }  \notag \\
&&-\hat{b}_{A}\left( 0\right) e^{-i\omega _{A}t}\int_{0}^{t}dt^{\prime
}\left( t-t^{\prime }\right) e^{i\omega _{A}t^{\prime }}\Psi \left(
-t^{\prime }\right) ,  \notag
\end{eqnarray}%
where%
\begin{equation}
\Psi \left( t\right) =\frac{g^{2}}{2N}\sum_{k=0}^{2N-1}e^{i\varepsilon _{k}t}
\end{equation}%
is the memory function\cite{Kurizki4}. Thus we can consider the decay of
atomic tunneling rate modified by an instantaneous projective measurement
with respect to the initial state of the total system. Unlike fermions,
there can be more than one boson in a site. Thus, in the following we will
investigate the decay rate of this system with respect to three different
initial states, and try to find the behavior difference between boson and
fermion.

First suppose the total system is initially prepared in a Fock state $\hat{b}%
_{A}^{\dagger }|0\rangle $ with only one atom in lattice site $A$. By $M$
successive instantaneous projective measurements into $\hat{b}_{A}^{\dagger
}|0\rangle $, the decay rate is of the form
\begin{equation}
\Gamma =\frac{g^{2}\tau }{4\pi N}\sum_{m=0}^{2N-1}sinc^{2}\left[ \left(
J\cos \frac{m\pi }{N}-\frac{\omega _{A}}{2}\right) \tau \right] \text{,}
\label{v-sb}
\end{equation}%
which is exactly the fermion tunneling rate with magnetic flux $\phi =0$. It
can be seen from Eq.(\ref{v-sb}) that the atomic tunneling rate only depends
on three parameters: the time interval $\tau $ between two successive
measurements, the number of lattice sites arranged on the ring, and the
on-site potential $\omega _{A}$ which is controlled by the laser intensity,
but only $\tau $ and $\omega _{A}$ can be adjusted experimentally. When $%
\tau $ is very small and approaches zero, the well-known quantum Zeno effect
occurs, and the system's evolution is frozen. For a finite number of sites,
when $\tau $ has a finite value, the quantum Zeno effect and anti-Zeno
effect can be switched by adjusting the laser intensity: the quantum Zeno
effect occurs when controllable variable $\omega _{A}=2J\cos \left( m\pi
/N\right) $, and the anti-Zeno effect occurs when on-site potential $\omega
_{A}\neq 2J\cos \left( m\pi /N\right) $ or $\left\vert \omega
_{A}\right\vert >2\left\vert J\right\vert $. Also for a finite number of
sites, when no measurement is performed, the system decays rapidly and the
atom never goes back to site $A$ when $\omega _{A}=2J\cos \left( m\pi
/N\right) $ for arbitrary $m$; when $\omega _{A}\neq 2J\cos \left( m\pi
/N\right) $ or $\left\vert \omega _{A}\right\vert >2\left\vert J\right\vert $%
, the system never evolved and the atom stay in site $A$ for ever. When the
number of sites $2N\rightarrow \infty $, the energy of the ring array become
continuous, and thus for a proper $\tau $, the switch between quantum Zeno
and anti-Zeno effect is determined by whether $\left\vert \omega
_{A}\right\vert $ is larger or smaller than $2\left\vert J\right\vert $.

Now we consider the case with one site containing particles more than one.
Assume the initial state of this total system is a number state $\left\vert
n_{A}\right\rangle $ with all $n$ particles in site $A$. After time $t$, the
probability for finding $\left\vert n_{A}\right\rangle $ is
\begin{equation}
p\left( t\right) =\frac{1}{n!}\left\vert \left\langle 0\right\vert \left[
\hat{b}_{A}\left( t\right) \right] ^{n}\left\vert n_{A}\right\rangle
\right\vert ^{2}\text{.}  \label{v-np}
\end{equation}%
By substituting Eq.(\ref{v-op}) into Eq.(\ref{v-np}), we find that after $M$
successive projective measurements into the initial state, the probability
modified by the measurements has the similar form to Eq.(\ref{IV-4})
\begin{equation}
p\left( t\right) =\exp \left[ -2nt\int_{0}^{t}dt^{\prime }G(\tau ,t^{\prime
})e^{-i\omega _{A}t^{\prime }}\Psi \left( -t^{\prime }\right) \right] \text{.%
}
\end{equation}%
Through defining the modulation function introduced in Eq.(\ref{IV-5}), in
the energy spectra, we find the atomic tunneling rate is $n$ times lager
than that of fermion
\begin{equation}
\Gamma =n\frac{g^{2}\tau }{4\pi N}\sum_{m=0}^{2N-1}sinc^{2}\left( J\cos
\frac{m\pi }{N}-\frac{\omega _{A}}{2}\right) \tau \text{.}  \label{V-0}
\end{equation}%
The value of Eq.(\ref{V-0}) is demanded by four external controllable
parameters: the time interval $\tau $, the number of sites placed on the
ring, the on-site potential $\omega _{A}$ and the total number $n$ of atoms
in the entire system. The new controllable element $n$ is added due to the
boson enhancement effect. When $n$ is large, the bosonic atoms have a strong
trend to leave site $A$. This just exhibits quantum statistic effect in
quantum measurement for the localization of boson system.

Except for the $n$ enhancing decay of the boson atomic tunneling, the
situation we discussed above is not surprising since they are very similar
to that of fermion. To show the special features of the boson tunneling
control, we consider the case with the initial state of this total system
prepared in a quasi-classical state - the coherent state $|\alpha
_{A}\rangle =D_{A}(\alpha )|0\rangle $, where%
\begin{equation}
D_{A}(\alpha )=e^{{\alpha \hat{b}_{A}^{\dagger }-\alpha ^{\ast }\hat{b}_{A}}}
\end{equation}%
is the displace operator. Like the Fock state listed above, this coherent
state is also unstable, and the atoms at site $A$ may tunnel to the array.
Once atoms are found in one site of the array, they will spread on the array
by resonant tunneling. Thus, it is difficult for all the atoms to go back to
site $A$. In order to keep all the atoms in their original state, a sequence
of measurements are performed, which project the entire system into $|\alpha
_{A}\rangle $. A measurement projects the system into the original state
with probability%
\begin{equation}
p\left( t\right) =\left\vert e^{-\left\vert \alpha \right\vert
^{2}/2}\left\langle 0\right\vert e^{\alpha ^{\ast }\hat{b}_{A}\left(
t\right) }\left\vert \alpha \right\rangle \right\vert ^{2}.  \label{v-pc}
\end{equation}

To calculate the explicit expression of the above probability, we define%
\begin{equation}
r\equiv \int_{0}^{t}dt^{\prime }\left( 1-\frac{t^{\prime }}{t}\right)
e^{-i\omega _{A}t^{\prime }}\Psi \left( -t^{\prime }\right) .  \label{v-dr}
\end{equation}%
As the evolution of $\hat{b}_{A}\left( t\right) $ is already obtained in Eq.(%
\ref{v-op}), we obtain the explicitly expression of probability
\begin{eqnarray}
p\left( t\right)  &=&e^{-\left\vert \alpha \right\vert ^{2}\left( \left\vert
\eta \left( t\right) \right\vert ^{2}+3-4\cos \left( \omega _{A}\tau \right)
\eta \left( t\right) \right) }\times   \notag \\
&&\exp \left[ -\sum_{m}\frac{\left\vert \alpha g\right\vert ^{2}\sin
^{2}[\left( \varepsilon _{m}-\omega _{A}\right) \frac{t}{2}]}{N\left(
\varepsilon _{m}-\omega _{A}\right) ^{2}}\right] ,
\end{eqnarray}%
where $\eta \left( t\right) =1-rt$. After $M$ successive projective
measurements, we find the atomic tunneling rate $\Gamma $ is modified as%
\begin{equation}
\Gamma =\frac{\left\vert \alpha \right\vert ^{2}}{\tau }\left[ \left\vert
\eta \left( t\right) \right\vert ^{2}+3-2\cos \left( \omega _{A}\tau \right)
\eta \left( t\right) -\pi r\tau ^{2}\right] .  \label{V-1}
\end{equation}%
Here the expression of $r$ is transformed into the following form through
Fourier transformation%
\begin{equation}
r=\frac{g^{2}\tau }{4\pi N}\sum_{m=0}^{2N-1}sinc^{2}\left[ J\cos \frac{m\pi
}{N}-\frac{\omega _{A}}{2}\right] \tau .  \label{V-2}
\end{equation}

In order to study the physical phenomena with $|\alpha _{A}\rangle $ as the
initial state,
%%%%%%%%%%%%%%%%%%%%%%%%%%%%%%%%%%%%%%%%%%%%%%%%%%%%%%%%%%%%%%%%%%%%
\begin{figure}[tbp]
\includegraphics[bb=35 195 550 651, width=8 cm,clip]{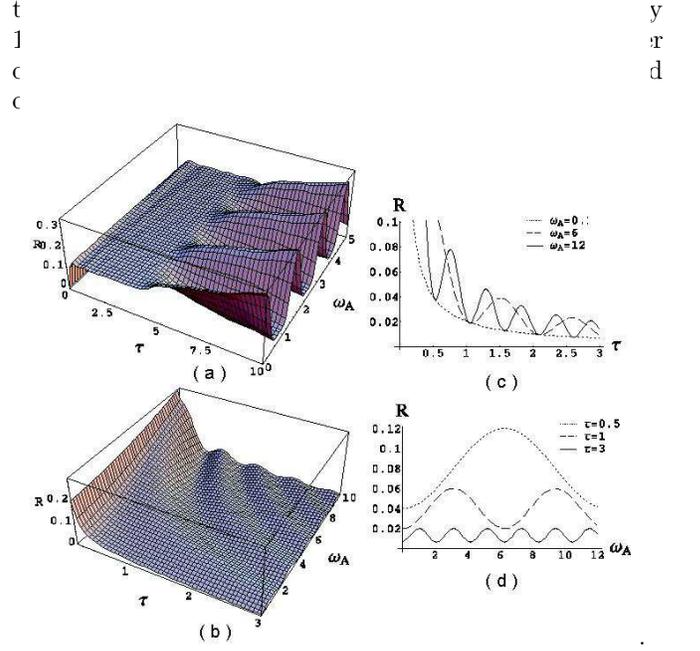} .
\caption{\textit{(Color on line) The behavior of the decay rate as a
function of $\protect\tau $ and $\protect\omega _{A}$ with the system
initial in a coherent state. (a) 3-D diagram for Fermi system with $%
J=5,g=2,N=20,\protect\phi =0$. (b) 3-D diagram for Boson system with $%
J=5,g=0.01,N=20,\protect\alpha =0.1$. (c) The cross sections of the 3-D
surface for $\protect\omega _{A}=0.1,6,12$. (d)The cross sections of the 3-D
surface for $\protect\tau =0.5,1,3$. It shows the tunneling rate can be
slightly modulated by the intensity of laser beam, but in the gross, $\Gamma
$ is a decreasing function of $\protect\tau $ for any $\protect\omega _{A}$}%
. }
\label{boson}
\end{figure}
%%%%%%%%%%%%%%%%%%%%%%%%%%%%%%%%%%%%%%%%%%%%%%%%%%%%%%%%%%%%%%%%%%%%
in Fig.\ref{boson}, we numerically plot the decay rate as a function of two
controllable external parameters $\tau $ and $\omega _{A}$. It shows that:
1) For a given on-site potential $\omega _{A}$, as $\tau \rightarrow 0$,
this unstable state decays rapidly. This phenomenon is totally different
from the fermion case, where the electron is frozen in its initial state. 2)
For any on-site potential $\omega _{A}$, the tunneling rate can be slightly
modulated by the intensity of laser beam, but in the gross, it is enhanced
as the measurement frequency $1/\tau $ increasing. However in fermi system,
the crossover of quantum Zeno and anti-Zeno effect can be controlled only by
modulation of on-site potential.

\section{\label{sec:summary}Summary}

In conclusion, we have investigated the quantum tunneling dynamics
for both fermion and boson systems in an experimentally accessible
engineered configuration respectively. In the case with electrons,
the tunneling rate modified by the projective measurements can be
controlled by the time interval between two successive measurements,
the electrochemical gate electrode and the magnetic flux. Our
results show that: 1) whatever the value of on-site potential
$\omega _{A}$ is, for vanishing time interval, the quantum Zeno
effect happens; 2) for $\omega _{A}$ off resonance with the energy
of the dot array, the quantum anti-Zeno effect occurs as the
measurement frequency increases; 3) for the on-site potential
$\omega _{A}$ resonating the energy of the dot array, we can inhibit
or accelerate the evolution of the electron by adjusting the
magnetic flux and the on-site potential. In the case of boson
system, generally, the time interval and the laser intensity control
the decay of the system. The boson system shows an enhanced decay
for quantum tunneling.

This work is supported by the NSFC with grant Nos. 90203018, 10474104 and
60433050, and NFRPC with Nos. 2001CB309310 and 2005CB724508. One (LZ) of the
authors also acknowledges the support of K. C. Wong Education Foundation,
Hong Kong.

\end{document}